\documentclass[adp,a4paper,fleqn%
]{w-art}
\usepackage{times,cite,w-thm}
\usepackage{graphicx}
\usepackage{amssymb,amsmath}
\newcommand{\unitmatrix}{1\hspace{-0.12cm}\mbox{I}}
\newcommand{\ket}[1]{{| #1 \rangle}\xspace}

\begin{document}
	\DOIsuffix{??}
	\Volume{??}
	\Month{??}
	\Year{2014}
	\pagespan{1}{}

	\title[Exact Ground State of Strongly Correlated Systems from SEMF]
	  {Exact Ground State of Strongly Correlated Electron Systems from Symmetry-Entangled Wave-Functions}

	\author[A. Lepr\'evost]{Alexandre Lepr\'evost\inst{1}}
	\author[O. Juillet ]{Olivier Juillet\inst{1}}
	\author[R. Fr\'esard]{Raymond Fr\'esard\inst{2}
	\footnote{Corresponding author\quad
			E-mail:~\textsf{Raymond.Fresard@ensicaen.fr},
			Phone: +33\,231\,45\,26\,09,
			Fax: +33\,231\,95\,16\,00}}
        \address[\inst{1}]{Laboratoire LPC Caen, ENSICAEN, Universit\'e de
          Caen, CNRS/IN2P3, 6 Boulevard Mar\'echal Juin, 14050 Caen CEDEX, France}
	\address[\inst{2}]{Laboratoire CRISMAT, UMR CNRS-ENSICAEN(ISMRA) 6508,
          6 Boulevard Mar\'echal Juin, 14050 Caen CEDEX, France}

	\Receiveddate{}  %for the publisher
	\Reviseddate{}
	\Accepteddate{}
	\Dateposted{}


	\begin{abstract}
		The four-site Hubbard model is considered from the exact
                diagonalisation and variational method points of view. It is
                shown that the exact ground-state can be recovered by a
                symmetry projected Slater determinant, irrespective of the
                interaction strength. This is in contrast to the Gutzwiller
                wave-function, which is calculated as well.
	\end{abstract}

	\maketitle

\section{Introduction}\label{sec:intro}

The intriguing properties of transition metal oxides have a long history,
perhaps starting with the pioneering work of de Boer and
Verwey on systems with partially filled and with completely filled
3d-bands\cite{deB37}. Since then, tremendous efforts have been devoted to the
study of transition metal oxides, especially in the form of perovskites
ABO$_3$, with A 
being, e.~g., La, Sr or a mixture of both of them, and B any transition
metal. As reviewed by Imada \textit{et.~al.}\cite{Ima98}, numerous phase
transitions have been discovered, especially towards magnetic and
superconducting states\cite{Koo67}, as well as the Mott 
metal-to-insulator transition. Arising in systems with partially filled bands,
it points towards the relevance of electronic correlations. Examples are
provided by RTiO$_3$ (see, e.g.,\cite{Tok93}) and RNiO$_3$ (see,
e.g., \cite{Med97} for a review), with R being a rare earth. Furthermore, the  
colossal magnetoresistance (see, e.g.,\cite{Hel93}) and large thermopower (see,
e.g.,\cite{Oht07}) have attracted considerable interest, too.

The transition metal oxides family is richer, and interest in systems
exhibiting application-oriented properties immensely grew  in recent
years. This is especially true for high-T$_c$ superconductors (see, e.g.,
\cite{Malozemoff05,Bed86}), transparent conducting oxides (see,
e.g.,\cite{Kawa97}), quantum criticality (see, e.g.,\cite{Buet10}), and high
capacitance heterostructures~\cite{Li11}, to 
quote a few. In addition, they also entail fascinating phenomena such as
superconductivity at the interface of two insulators \cite{Reyren06}, peculiar
magnetism in low dimensional systems \cite{Eyert08}, high temperature
ferromagnetism in vanadate superlattices \cite{Lud09}, all of them providing
a strong challenge to investigate these systems from the theory side. Yet, one
may fairly say that current theoretical approaches meet with severe
difficulties when studying the models which describe these systems. Indeed,
the tool which is best mastered (perturbation theory), badly fails when the
Coulomb interaction is sufficiently strong, thereby
calling for alternative approaches. 

The microscopic model for interaction driven properties which has received the
largest amount of attention is certainly the Hubbard model
\cite{Gut63,Hub63,Kan63}, especially after Anderson's proposal that the key
properties of high-T$_c$ superconductors are embodied in it or in the t-J
model \cite{And87}. The Hubbard model describes an interacting many-body
system which cannot be solved analytically, except for dimension
$d=1$\cite{Lie68}, or in the limit of large spacial
dimensions\cite{Met89,Mul89,Geo96}. 

In two dimensions, consensus has been reached at half-filling only: the
interaction strength drives a Mott transition\cite{Ass96} to an insulator with
long-ranged antiferromagnetic (AF) order \cite{Hir85}. The consequences of
(hole) doping remain controversial: The debate focuses on whether the
ground-state supports charge inhomogeneities or unconventional fermion-pair 
condensates and, if so, how their order parameters are intertwined with magnetic
properties. Even though numerous many-body techniques have been applied, 
only a partial answer could be  obtained. As reviewed in \cite{Ave11}, they
are, e.~g., cluster 
extensions \cite{Mai05} of the dynamical mean-field theory \cite{Geo96}, the 
two-particle self-consistent approximation\cite{AMT06}, Gutzwiller variational
schemes \cite{Ede07}, or slave-boson approaches \cite{Rac06,Sei04}. Standard 
quantum Monte Carlo simulations (QMC) are also restricted \cite{Whi89}
owing to the notorious sign problem that is particularly severe
for doped Hubbard models. 

An alternative approach with unrestricted symmetry projected wave-functions
has been recently introduced. This symmetry-entangled mean-field (SEMF) theory
is variational\cite{Jui13}, and has been 
shown to be exact for a two-site cluster, sharing this property with the
Gutzwiller wave-function (GWF). In addition, excellent agreement with exact
diagonalisation data on $4\times 4$ clusters has been obtained
\cite{suppl}. This then raises the question of whether the SEMF could be exact 
for clusters made of more than two sites, which provides the purpose of
this paper. It is organised as follows:  We
first summarise in Section~\ref{sec:ed} the exact diagonalisation procedure to
the calculation of the ground-state of the Hubbard model on a $2\times 2$
cluster which we determine. We then present the principles of the SEMF in
Section~\ref{sec:prsemf} and
work out the wave-function obtained after restoration of the spin rotational
invariance. For comparison, we also determine the Gutzwiller wave-function
as well as the Hartree-Fock 
wave-function. Finally, in Section~\ref{sec:exa} we calculate the SEMF wave-function obtained
after spin and total momentum projection, and we show that the exact
ground-state energy is recovered for arbitrary interaction strength. Our work is summarised in Section~\ref{sec:conclusion}.

\section{Exact diagonalisation for the ground-state}\label{sec:ed}

\begin{figure}[t!]
	\centering
	\includegraphics[width=0.4\columnwidth]{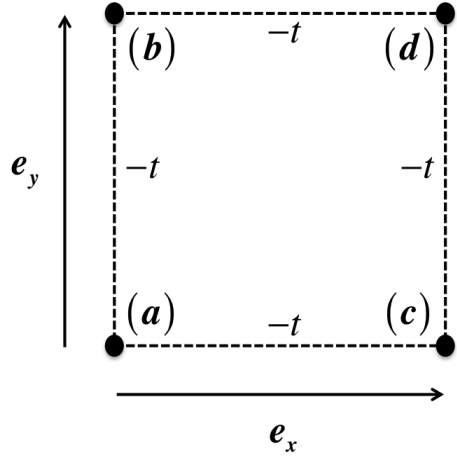}
	\caption{Labelling of the sites for the Hubbard model on the $2\times
          2$ cluster. 
	}
	\label{fig:lab}
\end{figure}

In this work we consider the Hubbard Model on four sites $i\in \{a,b,c,d\}$ as
depicted in Fig.~\ref{fig:lab}. We write the Hamiltonian in the form  
\begin{equation}\label{eq:H}
{\mathcal H} = -\frac{t}{2} \sum_{<i,j>,\sigma} c^{\dagger}_{i,\sigma}
c^{\phantom{\dagger}}_{j,\sigma} + U  \sum_i n^{\phantom{\dagger}}_{i,\uparrow} n^{\phantom{\dagger}}_{i\downarrow}
\end{equation}
Note that the unusual factor $\frac{1}{2}$ for the hopping amplitude is simply
introduced to avoid double counting following from the periodic boundary
conditions (PBC) we use. This Hamiltonian
is invariant under both continuous and discrete symmetry operations. They
involve the $SU(2)$ spin rotational symmetry, as well as translational
symmetry and $C_{4v}$ lattice transformations.  In addition, at
half-filling, the Hamiltonian
Eq.~(\ref{eq:H}) possesses $SU(2)$ charge rotational symmetry. As it turns out
that this symmetry is not needed to obtain the exact ground-state in the SEMF
approach, it will be discarded.

According to previous studies \cite{Lie89,Mor90}
the ground-state is 
characterised by zero total spin, zero total momentum and d-wave symmetry.  A
convenient basis to the calculation of the ground-state may be found starting
from  a state with two doubly occupied sites that is symmetrised according
to the above quantum numbers. The other basis states are obtained by
repeatedly applying the hopping operator. Thus, we end with a three-dimensional
subspace spanned by the following vectors:
\begin{eqnarray}\label{eq:basis}
\ket{1}&=& \frac{1}{2} 
\left( \Delta^{\dagger}_{a} - \Delta^{\dagger}_{d} \right)
\left(\Delta^{\dagger}_{b} -  \Delta^{\dagger}_{c} \right)
\ket{0}\nonumber \\
\ket{2}&=& \frac{1}{4} \left[ 
\left(\Delta^{\dagger}_{a} + \Delta^{\dagger}_{b}\right) 
\left( c^{\dagger}_{c \uparrow} c^{\dagger}_{d \downarrow}
 - c^{\dagger}_{c \downarrow} c^{\dagger}_{d \uparrow} \right) - 
\left(\Delta^{\dagger}_{a} + \Delta^{\dagger}_{c}\right) 
\left( c^{\dagger}_{b \uparrow} c^{\dagger}_{d \downarrow}
 - c^{\dagger}_{b \downarrow} c^{\dagger}_{d \uparrow} \right) \right. \nonumber \\
&&\left. - 
\left(\Delta^{\dagger}_{b} + \Delta^{\dagger}_{d}\right) 
\left( c^{\dagger}_{a \uparrow} c^{\dagger}_{c \downarrow}
 - c^{\dagger}_{a \downarrow} c^{\dagger}_{c \uparrow} \right) +
\left(\Delta^{\dagger}_{c} + \Delta^{\dagger}_{d}\right) 
\left( c^{\dagger}_{a \uparrow} c^{\dagger}_{b \downarrow}
 - c^{\dagger}_{a \downarrow} c^{\dagger}_{b \uparrow} \right)
\right]
\ket{0}\nonumber \\
\ket{3}&=& \frac{1}{2\sqrt{3}} \left[ 
c^{\dagger}_{a \uparrow} c^{\dagger}_{b \uparrow} c^{\dagger}_{c \downarrow} c^{\dagger}_{d \downarrow}
+c^{\dagger}_{a \downarrow} c^{\dagger}_{b \uparrow} c^{\dagger}_{c \downarrow} c^{\dagger}_{d \uparrow}
+c^{\dagger}_{a \downarrow} c^{\dagger}_{b \downarrow} c^{\dagger}_{c \uparrow} c^{\dagger}_{d \uparrow}
+c^{\dagger}_{a \uparrow} c^{\dagger}_{b \downarrow} c^{\dagger}_{c \uparrow}
c^{\dagger}_{d \downarrow} \right. \nonumber \\
&&\left. -2\left(
c^{\dagger}_{a \uparrow} c^{\dagger}_{b \downarrow} c^{\dagger}_{c \downarrow}
c^{\dagger}_{d \uparrow} +
c^{\dagger}_{a \downarrow} c^{\dagger}_{b \uparrow} c^{\dagger}_{c \uparrow}
c^{\dagger}_{d \downarrow} \right) \right]
\ket{0}
\end{eqnarray}
where we introduced the short-hand notation $ \Delta^{\dagger}_{i} \equiv
c^{\dagger}_{i \uparrow} c^{\dagger}_{i \downarrow}$. In this basis the
Hamiltonian matrix reads:
\begin{equation}\label{eq:Hmatr}
H = \left(\begin{array}{ccc}
2U&2t&0\\
2t&U&-2\sqrt{3}t\\
0&-2\sqrt{3}t&0
\end{array}\right)
\end{equation}
In order to determine the eigenvalues $E_k$ it is convenient to write: 
\begin{equation}\label{eq:defzeta}
E \equiv U -4 t \zeta.
\end{equation}
$\zeta$ then satisfies to a cubic equation in depressed form:
\begin{equation}\label{eq:eigszeta}
16 t^3 \zeta^3 - \zeta(16 t^2 + U^2) t - 2 t^2 U = 0.
\end{equation}
Eq.~(\ref{eq:eigszeta}) may be solved using Cardano's formula, and the
eigenvalues of the Hamiltonian matrix Eq.~(\ref{eq:Hmatr}) finally read: 
\begin{equation}\label{eq:sols}
E_k = U -2 \sqrt{\frac{16 t^2 + U^2}{3}} \cos{\left(\frac{\beta - 2 k
      \pi}{3}\right)}
\end{equation}
with $k=0,~1,~2$, and
\begin{equation}\label{eq:solbet}
\cos{\beta} = 4 t^2 U \left(\frac{3}{16 t^2 + U^2}\right)^{\frac{3}{2}}.
\end{equation}
The ground-state corresponds to $k=0$. 

The Hamiltonian Eq.~(\ref{eq:H}) also
corresponds to the Hubbard Model on a four-site chain with PBC. Note that the
d-wave character of the ground-state on the $2\times 2$ cluster maps onto a
total momentum $K=\pi$ for the chain. Then, the solution
Eqs.~(\ref{eq:sols}-\ref{eq:solbet}) reproduces the results obtained in the 1D
case \cite{Cuo96,Sch02}. 

\section{Principles of the SEMF approach for the $2\times 2$ cluster}\label{sec:prsemf}

Symmetry breaking wave-functions with assumed magnetic, charge or
superconducting orders are usually considered in variational treatments of the
Hubbard model \cite{varia}. However, on finite-size clusters, Hamiltonian
symmetries must be restored by quantum fluctuations and substantial energy
improvements can be obtained by quantum number projection on
top of these states. An illustration with the Gutzwiller wave-function can
be found in \cite{Ima08}. Indeed, symmetry restoration leads to coherent
superpositions of symmetry related states that induce correlations. The SEMF
approach follows such a strategy to approximate low-lying eigenstates of the
Hubbard model. Up to now, the method works at the Hartree-Fock level and gives
the optimal Slater determinant minimising the energy \textit{after} symmetry
projection. First attempts for the spectroscopy of Hubbard chains \cite{Schm}
and square 
clusters up to 36 sites have been performed with reliable results
\cite{Scus}. By reformulating the stationarity of the projected energy as a
mean-field like equation, SEMF simulations on larger cells and with all
symmetries of the Hubbard model have revealed an intriguing interplay of spin,
charge and pairing correlations in the hole doped region \cite{Jui13}. 

\subsection{Restoration of spin-rotational invariance}

One of the most attractive features of the SEMF approximation relies on the
ability to perform an unbiased energy minimisation, \textit{i.e.} to consider
totally unrestricted Slater determinants. However, we here focus  on analytical
SEMF solutions and thus follow conventional calculations with projected
wave-functions by constraining the variational subspace to exhibit a relevant
order. At half-filling, an antiferromagnetic Slater determinant
$\ket{\Phi_{\rm ref}}$ is considered. For the $2\times 2$ cluster, we assume a
positive magnetisation $m$ on sites a and d, and the opposite magnetisation on
sites b and c. In each spin sector $\sigma$, the two occupied orbitals are
simply obtained as the lowest energy eigenstates of the effective one-body
Hamiltonian 
\begin{equation}\label{eq:hmf}
h_{\sigma}^{\rm ref}[m] = \left(\begin{array}{cccc}
\frac{U}{2}(1-\sigma m)&-t&-t&0\\
-t&\frac{U}{2}(1+\sigma m)&0&-t\\
-t&0&\frac{U}{2}(1+\sigma m)&-t\\
0&-t&-t&\frac{U}{2}(1-\sigma m)
\end{array}\right).
\end{equation}
For our purpose, it is convenient to introduce
\begin{eqnarray}\label{eq:msca}
\tilde{m}&=&\frac{Um}{2t} \nonumber \\
\cos{(2\varphi)}&=&\frac{\tilde{m}}{\sqrt{4+\tilde{m}^2}} \nonumber \\
\sin{(2\varphi)}&=&\frac{2}{\sqrt{4+\tilde{m}^2}}
\end{eqnarray}
in terms of which the four-electron state $\ket{\Phi_{\rm ref}}$ reads:
\begin{equation}\label{eq:phiref}
\ket{\Phi_{\rm ref}} = c^{\dagger}_{\phi_1 \uparrow} c^{\dagger}_{\phi_2 \uparrow} c^{\dagger}_{\phi_3 \downarrow} c^{\dagger}_{\phi_4 \downarrow}\ket{0} 
\end{equation}
with
\begin{eqnarray}\label{eq:eigs}
|\phi_1\rangle &=& \frac{1}{\sqrt{2}} \left( |a \rangle - |d \rangle \right) \nonumber \\
|\phi_2\rangle&=& \frac{1}{\sqrt{2}} \left( \cos{(\varphi)} |a \rangle + 
\sin{(\varphi)} |b \rangle + \sin{(\varphi)} |c \rangle + 
\cos{(\varphi)} |d \rangle \right)\nonumber \\
|\phi_3\rangle&=& \frac{1}{\sqrt{2}} \left( |b \rangle - |c \rangle \right)\nonumber \\
|\phi_4\rangle&=& \frac{1}{\sqrt{2}} \left( \sin{(\varphi)}  |a \rangle +
  \cos{(\varphi)} |b \rangle + \cos{(\varphi)} |c \rangle + 
\sin{(\varphi)} |d \rangle \right)
\end{eqnarray}
In SEMF, one introduces a symmetry adapted mean-field state
$\ket{\Psi} = {\mathcal P}^{(\Gamma)} \ket{\Phi_{\rm ref}}$ where ${\mathcal P}^{(\Gamma)}$ is a projector on the
subspace with quantum numbers $\Gamma$. We first limit ourselves to the
restoration of spin rotational invariance for which the singlet projection can
be achieved by \cite{Ham62}
\begin{equation}\label{eq:pseq0}
{\mathcal P}^{(S=0)} = \frac{1}{8 \pi^2} \int_0^{2\pi} d\alpha  \int_0^{\pi}
d\beta \sin{\beta} \int_0^{2\pi} d\gamma R(\alpha,\beta,\gamma)
\end{equation}
where $R(\alpha,\beta,\gamma) = e^{−i\alpha S_z} e^{−i\beta S_y}
e^{−i\gamma S_z}$   is the Euler angles $(\alpha,\beta,\gamma)$
parameterisation of rotations (with $S$ the total spin observable). The
unnormalised $S=0$ component of the AF state is then easily obtained and may be written as a linear combination of the basis vectors $\ket{1}$, 
$\ket{2}$, $\ket{3}$ Eq.~(\ref{eq:basis}) spanning the subspace of the exact
ground-state : 
\begin{equation}\label{neweq2}
\ket{\Psi} = \sin^2{(\varphi)}\ket{1} -  \sin{(2\varphi)}\ket{2} - 
\frac{1+\cos^2{(\varphi)}}{\sqrt{3}} \ket{3}
\end{equation}
Indeed, this result reflects unbroken symmetries of the AF reference state. 
For instance, one can immediately check that $\ket{\Phi_{\rm ref}}$ is
invariant under a spin-rotation around the y-axis by an angle $\pi$
combined with a translation $T_y$ by one lattice spacing along the
y-direction. Therefore, after performing all spin rotations and integrating
over the Euler angles, the resulting $S=0$ vector is simultaneously
translational invariant: 
\begin{eqnarray}\label{neweq3}
T_y \ket{\Psi} &=& \frac{1}{4 \pi} \int_0^{2\pi} d\alpha e^{−i\alpha S_z}
\int_0^{\pi} d\beta \sin{\beta} e^{−i\beta S_y} T_y 
\ket{\Phi_{\rm ref}}\nonumber \\
&=& \frac{1}{4 \pi} \int_0^{2\pi} d\alpha' e^{-i(\alpha' -2 \pi) S_z} 
\int_0^{\pi} d\beta' \sin{(\pi-\beta')} e^{-i\beta' S_y} 
e^{−i\pi S_y} T_y \ket{\Phi_{\rm ref}}\nonumber \\
&=& e^{2 i \pi S_z} {\mathcal P}^{(S=0)}\ket{\Phi_{\rm ref}}~=~ \ket{\Psi}
\end{eqnarray}
where we used the hermiticity of the projector Eq.~(\ref{eq:pseq0}). With the
help of the Hamiltonian matrix Eq.~(\ref{eq:Hmatr}) in the subspace $\ket{1}$,
$\ket{2}$, and $\ket{3}$ , one is left with the following average energy
$E^{(S=0)}$ in the SEMF state $\ket{\Psi}$: 

\begin{equation}\label{eq:Esemf}
\langle {\mathcal H} \rangle_{{\mathcal P}^{(S=0)} \Phi_{\rm ref}}= \frac
{\frac{3}{8}U\left(5 -4 \cos{(2 \varphi)} - \cos{(4 \varphi)} \right) - 
12 t \sin{(2 \varphi)}}
{2+\sin^2{(2 \varphi)}}
\end{equation}
\begin{figure}[b!]
	\centering
	\includegraphics[width=0.5\columnwidth,angle=-90]{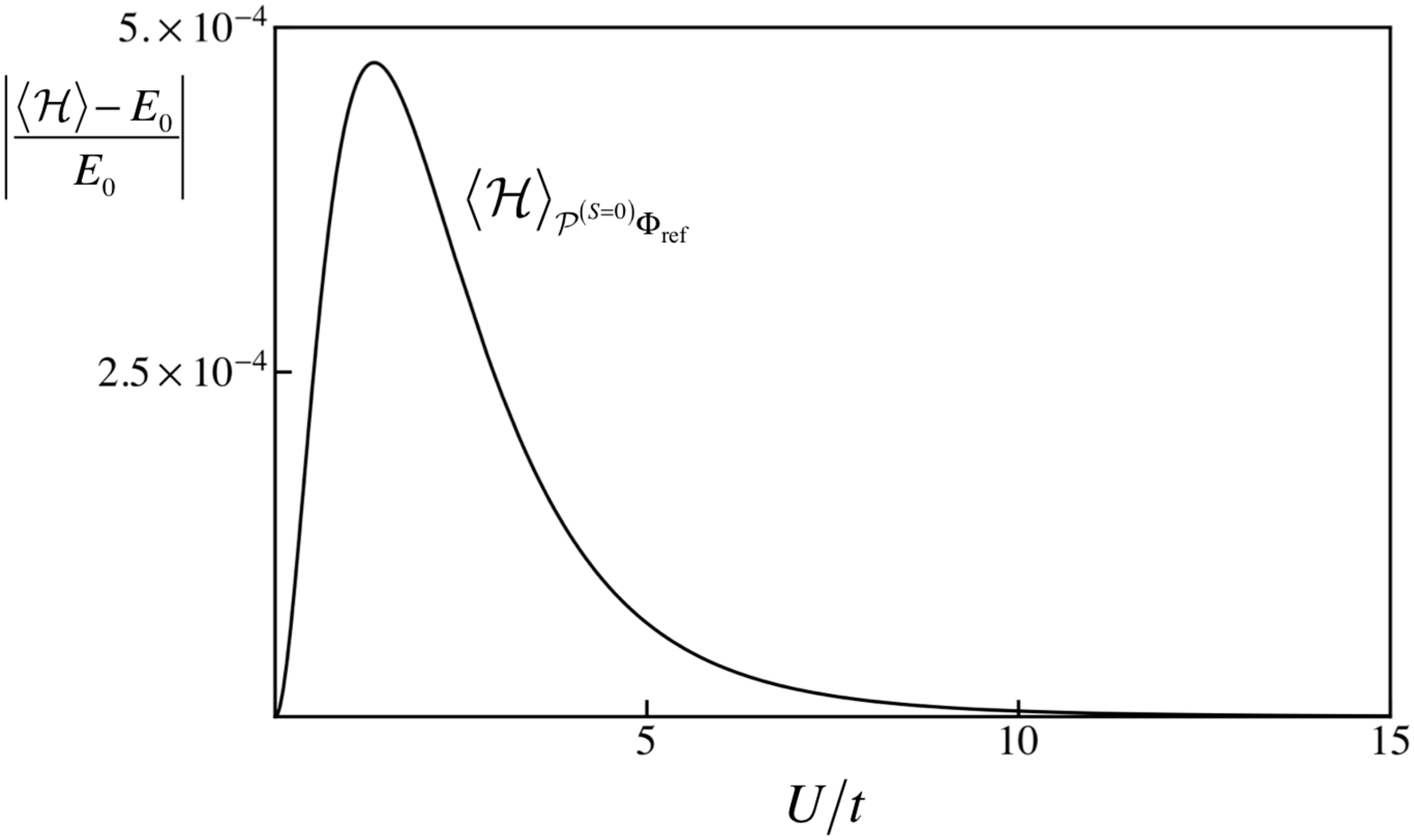}
	\caption{Ground-state energy from spin singlet projected SEMF. 
	}
	\label{fig:ensemf}
\end{figure}
Its minimisation with respect to $\varphi$ yields a cubic equation for the dimensionless staggered 
magnetisation $\tilde{m}$:
\begin{equation}\label{eq:spm}
2 \tilde{m} t (2 +\tilde{m}^2) - U (3 +\tilde{m}^2) =0
\end{equation}
Comparison of the resulting energy to the exact solution is performed in
Fig.~\ref{fig:ensemf}. The agreement is obviously excellent for any on-site
interaction, but the SEMF approach limited to the spin singlet projection is
not exact.

\subsection{Comparison with the conventional Gutzwiller wave-function}

Let us now proceed to the usual variational calculation with the Gutzwiller
projector 
\begin{equation}\label{eq:Gproj}
{\mathcal P}_G = \Pi_i\left(1+(g-1)
  n^{\phantom{\dagger}}_{i,\uparrow}
  n^{\phantom{\dagger}}_{i\downarrow}\right)
\end{equation}
to reduce the weights of configurations with double occupancy in the AF
reference state. Thus, one aims at minimising :
\begin{equation}\label{eq:GutzE}
\langle {\mathcal H} \rangle (\varphi,g) \equiv \frac{
\langle \Phi_{\rm ref}|{\mathcal P}_G {\mathcal H}{\mathcal P}_G | 
\Phi_{\rm ref}\rangle}
{\langle \Phi_{\rm ref}|{\mathcal P}_G {\mathcal P}_G | \Phi_{\rm ref}\rangle}
\end{equation}
with respect to $\varphi$ and $g$. While the evaluation of the norm of the
projected wave-function and the expectation value of the interaction energy is
straightforward, the one of the kinetic energy is more tedious. Yet,
symmetries of the AF background greatly simplify the calculation since all
hopping contributions are equal. For instance, from the invariance of 
$\ket{\Phi_{\rm ref}}$ under the product $U_y = T_y R(0,\pi,0)$, one has: 

\begin{eqnarray}\label{neweq4}
\langle \Phi_{\rm ref} |{\mathcal P}_G c^{\dagger}_{a \uparrow} 
c^{\phantom{\dagger}}_{c \uparrow} {\mathcal P}_G | \Phi_{\rm ref} \rangle &=&
\langle \Phi_{\rm ref} |{\mathcal P}_G U^{\dagger}_y c^{\dagger}_{a \uparrow} 
c^{\phantom{\dagger}}_{c \uparrow}U_y {\mathcal P}_G | \Phi_{\rm ref}\rangle
\nonumber \\
&=& \langle \Phi_{\rm ref} |{\mathcal P}_G c^{\dagger}_{b \downarrow} 
c^{\phantom{\dagger}}_{d \downarrow} {\mathcal P}_G | \Phi_{\rm ref}\rangle
\end{eqnarray}
where we have used the invariance of the Gutzwiller operator under symmetry transformations of the Hamiltonian. The average energy (12) is finally obtained as:
\begin{figure}[b!]
	\centering
	\includegraphics[width=0.5\columnwidth,angle=-90]{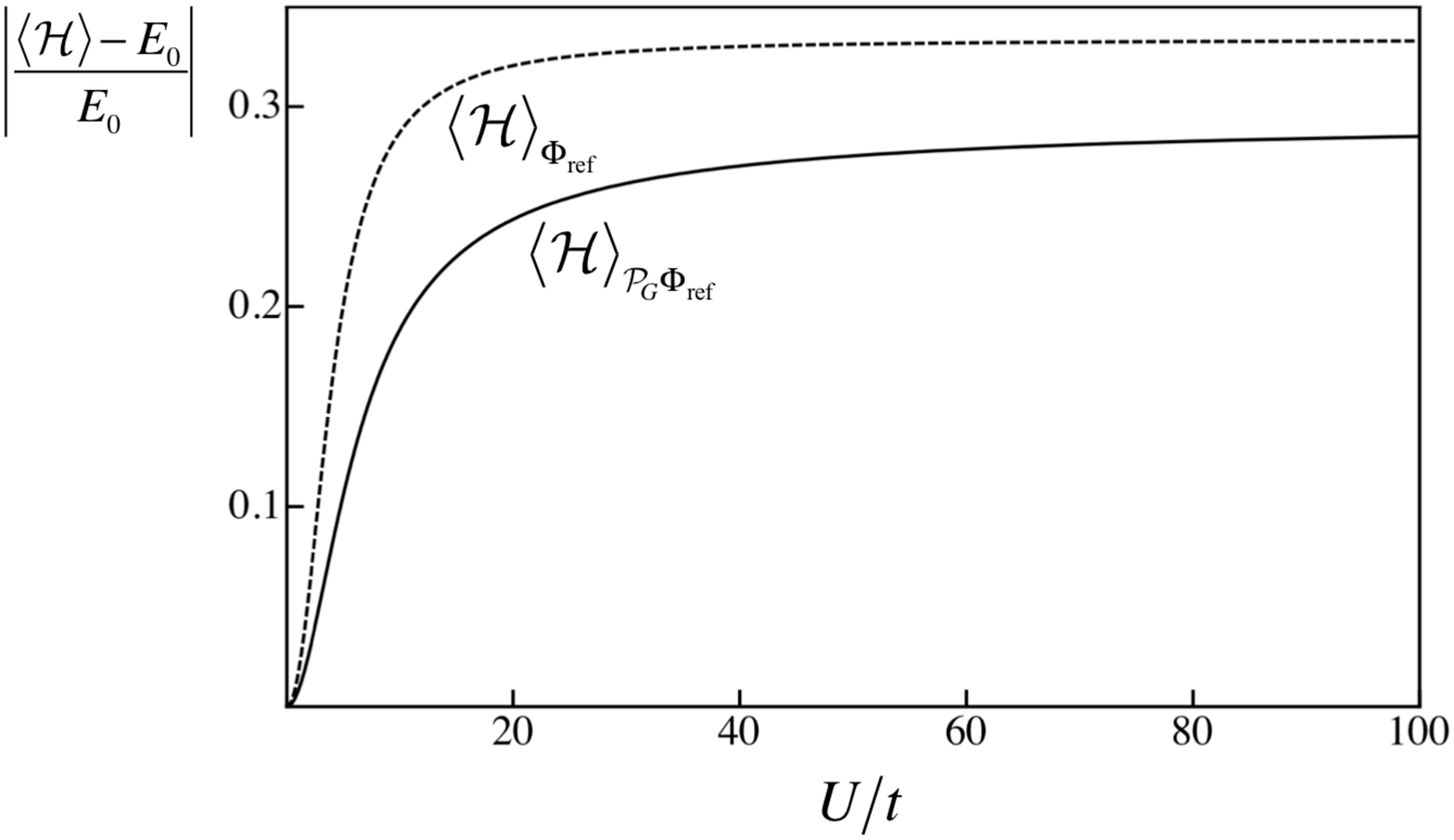}
	\caption{Ground-state energy from antiferromagnetic Gutzwiller
          wave-function (full line) and in mean-field theory (dashed line),
          relative to the exact one. 
	}
	\label{fig:enmf}
\end{figure}

\begin{equation}\label{eq:EG}
\langle {\mathcal H} \rangle_{{\mathcal P}_G \Phi_{\rm ref}} =
\frac{\left[-8tg\sin{(\varphi)} \cos{(\varphi)}+ 2 U g^2 \sin^2{(\varphi)} \right] 
\left[(1+g)^2 \sin^2{(\varphi)} + 4
      \cos^2{(\varphi)}\right]}
{4 \cos^4{(\varphi)} + 8 g^2 \cos^2{(\varphi)} \sin^2{(\varphi)} +(1+g^2)^2 \sin^4{(\varphi)}}
\end{equation}
Minimising Eq.~(\ref{eq:EG}) with respect to $g$ and $\varphi$ reveals that
the relative energy difference with the exact solution increases with $U$ and
saturates to $28\%$ for $U\rightarrow\infty$ as shown in
Fig.~\ref{fig:enmf}. Nevertheless, the Gutzwiller projection improves the
simple Hartree-Fock approximation which is recovered by imposing the staggered
magnetisation $m$ to solve the self-consistency condition 
$m=\langle n_{a \uparrow} - n_{a \downarrow} \rangle_{ \Phi_{\rm ref}}$. This
relation is equivalent to require a zero derivative of 
$\langle {\mathcal H}  \rangle_{ \Phi_{\rm ref}}$ with respect to $m$. The
energy obtained with such a mean-field solution is also displayed in
Fig.~\ref{fig:enmf} and gives a relative error of $33\%$ in the strong
coupling limit of the Hubbard model.

\section{The exact ground-state as a symmetry projected
  wave-function}\label{sec:exa} 
\begin{figure}[b!]
	\centering
	\includegraphics[width=0.4\columnwidth]{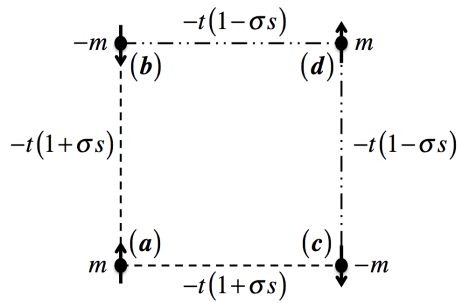}
	\caption{Mean-field like scheme defining the reference Slater
          determinant for the SEMF approach with spin and total momentum
          projection. 
	}
	\label{fig:ts}
\end{figure}

In the spirit of the SEMF methodology, breaking symmetries in the underlying
reference state followed by their restoration is a way to recover the small
part of the correlation energy that cannot be reached by the only projection
onto the spin-singlet subspace. We consider here a scheme obtained by adding a
bond-spin contribution to an antiferromagnetic order, as depicted in
Fig.~\ref{fig:ts}. As 
a result, the Slater determinant $\ket{\Phi_{\rm ref}}$ is built from the lowest
energy eigenstates of the mean-field like Hamiltonian: 
\begin{equation}\label{eq:hmfex}
h_{\sigma}^{\rm ref}[m,s] = \left(\begin{array}{cccc}
\frac{U}{2}(1-\sigma m)&-t(1+\sigma s)&-t(1+\sigma s)&0\\
-t(1+\sigma s)&\frac{U}{2}(1+\sigma m)&0&-t(1-\sigma s)\\
-t(1+\sigma s)&0&\frac{U}{2}(1+\sigma m)&-t(1-\sigma s)\\
0&-t(1-\sigma s)&-t(1-\sigma s)&\frac{U}{2}(1-\sigma m)
\end{array}\right).
\end{equation}
The SEMF variational ansatz then results from the projection on zero-spin and total momentum 
${\bf K}=(0,0)$:
\begin{equation}\label{eq:projps}
\ket{\Psi} = {\mathcal P}^{({\bf K}=0)} {\mathcal P}^{(S=0)} \ket{\Phi_{\rm ref}}
\end{equation}
where $P^{({\bf K}=0)}= \frac{1}{4}\left(\unitmatrix+T_x+T_y+T_{x+y} \right)$ ensures the invariance under translations $T_x$, $T_y$, $T_{x+y}$ by one lattice 
spacing along the directions ${\bf e_x}$, ${\bf e_y}$, ${\bf e_x} + {\bf e_y}$ \cite{Ham62}. Through similar 
steps as for 
spin projection, the  SEMF energy is given by 
\begin{eqnarray}\label{eq:semfmf}
E^{(S=0,{\bf K}=0)} &=& \frac{3(1+s^2)}{2} \times  \nonumber \\
%&&\frac
%{-8t\left(\tilde{m} s^2 + (2+s^2)\sqrt{4(1+s^2)+\tilde{m}^2} \right)
%+U(1+s^2)\left( 6+ \tilde{m}^2 + 2s^2 - \tilde{m}\sqrt{4(1+s^2)+\tilde{m}^2}\right) ]}
%{\tilde{m} s^2 \sqrt{4(1+s^2)+\tilde{m}^2} +2(1+s^2)(6+6s^2+s^4) + 
%\tilde{m}^2 (2+3s^2+2s^4)}\nonumber
&&\frac
{-8t\left(\tilde{m} s^2 + (2+s^2)A \right)
+U(1+s^2)\left( 6+ \tilde{m}^2 + 2s^2 - \tilde{m}A\right) ]}
{\tilde{m} s^2 A +2(1+s^2)(6+6s^2+s^4) + 
\tilde{m}^2 (2+3s^2+2s^4)}
\end{eqnarray}
with $A \equiv \sqrt{4(1+s^2)+\tilde{m}^2}$.  In order to solve analytically
the minimum equations is is convenient to introduce new variables, $\xi$ and
$\eta$. They are defined as:
\begin{eqnarray}\label{eq:xieta}
\xi&\equiv& \frac{2(2+s^2) + \tilde{m} \left(\tilde{m} 
+ \sqrt{4(1+s^2)+\tilde{m}^2} \right)}
{2\left(\tilde{m} + \sqrt{4(1+s^2)+\tilde{m}^2} \right)}
\nonumber \\
\eta&\equiv& \frac{2}{\sqrt{4(1+s^2)+\tilde{m}^2} -\tilde{m}},
\end{eqnarray}
and allow to cast the energy (\ref{eq:semfmf}) in the form
\begin{equation}\label{eq:exieta}
E^{(S=0,{\bf K}=0)} = \frac{3}{2} ~~ \frac{-16 \xi\eta^2t + U (1+2 \eta^2) }
{4 \xi^2\eta^2 + 3 \eta^2 - 2 \xi\eta +1 }.
\end{equation}
Minimisation with respect to $\xi$ and $\eta$ yields:
\begin{eqnarray}\label{eq:minxieta}
\xi^2 4 \eta (4 \eta t - U) -\xi\left((2\eta^2-1)U + 16\eta\right)&=&U\eta 
\nonumber \\
\xi^2 32 \eta^3 t - 4 \xi U \eta(2 \eta^2 +1 )&=& 8 \eta t(3 \eta^2+1) -U(2 \eta^2+1).
\end{eqnarray}
Eq.~(\ref{eq:minxieta}) may be viewed as a linear system of equations in $\xi$
and $\xi^2$ that is easily solved to 
express these variables in terms of $\eta$, \textit{i.e.} $\xi = f(\eta),
\xi^2 = g(\eta)$. Therefore, 
the relation $g(\eta) -f(\eta)^2=0$ has to be satisfied, which can be
factorised into  
\begin{eqnarray}\label{eq:eqb}
\left[ 8 \eta^3 t^2 - 6 \eta^2 t U + \eta (U^2-8t^2) + t U \right] &\times& \nonumber \\
\left[ 4  \eta^4(48 t^2 + U^2) - 16 \eta^3 t U + 4 \eta^2 (16 t^2 + U^2) - 8 
\eta t U + U^2\right] &=& 0 
\end{eqnarray}
The quartic factor 
$Q(\eta)$ as a function 
of $\eta>0$ is strictly positive for any interaction strength. Indeed, $Q(0)=U^2$ and $Q$ goes to 
infinity 
with $\eta$ which proves the result in the case of a monotonic evolution. Otherwise, the value 
$Q(\eta_0)$ at an extremum point $\eta_0$ can be obtained from the remainder of the Euclidean 
division of the polynomial $Q$ by its derivative :
\begin{equation}\label{eq:rest}
Q(\eta_{0}) = \frac{2 \eta_0^2(768 t^4 + 58 t^2 U^2 + U^4) -4 \eta_0 t U
(64 t^2 + U^2) + U^2 ( 46 t^2 + U^2)}{ 48 t^2 + U^2}
\end{equation}
Since the discriminant of the quadratic numerator is strictly negative, $Q(\eta_0)>0$ and so 
$Q(\eta)>0$ even for non-monotonic behaviour. Finally, the SEMF energy for $S=0$ and ${\bf K}=(0,0)$
is minimised if and only if $\eta$ is a root of the cubic factor in Eq.~(\ref{eq:eqb}). 
In this case, the difference $\eta-\xi = \eta-f(\eta)$ simplifies to the
interaction, up to a constant: 
\begin{equation}\label{neweq6}
\eta-\xi = \frac{U}{4t}
\end{equation}
Thus, the cubic equation for $\eta$ turns into a similar equation in terms of $\xi$ :
\begin{equation}\label{eq:cubxi}
16 t^2 \xi^3 -  \xi (16 t^2 + U^2) - 2 t U =0
\end{equation}
This relation must be satisfied for the projected energy Eq.~(\ref{eq:exieta})
to reach its minimum, given by  
\begin{equation}\label{neweq7}
E^{(s=0,{\bf K}=0)}_{\rm min} = U -4 \xi t
\end{equation}
We recover exactly Eq.~(\ref{eq:defzeta}) and Eq.~(\ref{eq:eigszeta}) obtained in Section~\ref{sec:ed} after direct
diagonalisation of the Hamiltonian matrix. Note that the result is valid for
arbitrary interaction strength. This SEMF derivation provides an alternative
view of the exact ground-state in terms of a one-parameter symmetry restored
Slater determinant. 
\section{Conclusion}\label{sec:conclusion}
In this work we showed analytically that conventional Hartree-Fock
approximations can be greatly improved to account for strong electronic
correlations as long as the variational ansatz is enhanced by symmetry
projections. For the $2\times 2$ cluster, we demonstrated that restoring the
spin-rotational invariance leads to an almost exact description. Furthermore,
we established that the more symmetry breakings in the underlying Slater
determinant followed by their restoration, the better the SEMF approach will
be. Specifically, spin-singlet and total momentum are sufficient to recover
the exact ground-state for any interaction strength. Combined with previous
numerical results on larger cells \cite{Jui13}, the present work highlights
the SEMF method as a reliable starting point to elucidate correlations that
spontaneously emerge from the Hubbard model at low energy. 
\begin{acknowledgement}
	This work was supported by 
        the R\'egion Basse-Normandie, and by the Minist\`ere de la Recherche.
\end{acknowledgement}

\end{document}